\documentstyle[aps,prb,epsf,multicol]{revtex}
\begin{document}

%\draft command makes pacs numbers print
\draft

\newcommand{\mybeginwide}{
    \end{multicols}\widetext
    \vspace*{-0.2truein}\noindent
    \hrulefill\hspace*{3.6truein}
}
\newcommand{\myendwide}{
    \hspace*{3.6truein}\noindent\hrulefill
    \begin{multicols}{2}\narrowtext\noindent
}

\title{Shot noise measurements in NS junctions and the semiclassical 
theory}
\author{X. Jehl\cite{present} and M. Sanquer}
\address{DRFMC-SPSMS, CEA-Grenoble, 38054 Grenoble cedex9, France.}
\date\today
\maketitle
\bigskip

\begin{abstract}

We present a new analysis of shot noise measurements in normal
metal-superconductor (NS) junctions [X. Jehl {\it et al.}, Nature {\bf
405}, 50 (2000)], based on a recent semiclassical theory.  The first
calculations at zero temperature assuming quantum coherence predicted shot
noise in NS contacts to be doubled with respect to normal contacts.  The
semiclassical approach gives the first opportunity to compare data and
theory quantitatively at finite voltage and temperature.  The doubling of
shot noise is predicted up to the superconducting gap, as already observed,
confirming that phase coherence is not necessary.  An excellent agreement
is also found above the gap where the noise follows the normal case.

\end{abstract}

\pacs{72.70.+m, 74.70.+k, 74.80.Fp}

\begin{multicols}{2}
\narrowtext

Noise properties of small NS junctions have attracted much attention in the
last years because they reveal features that are not accessible by linear
conductance measurements, and stimulate advances and confrontation of
different theoretical approaches.  Most of the peculiar features associated
with the so-called proximity effect in an NS bilayer were extensively
studied for more than 30 years \cite{deutscher}.  More recently, the
reentrance effect was discovered at very low temperatures in coherent
hybrid nanostructures, and the study of proximity effects at the
microscopic level revealed Andreev reflections as the main transport
mechanism through the interface \cite{courtois}.  Shot noise brings new
information because of its direct dependence on the carriers charge or the
interactions they experience, and already led to remarkable results
\cite{blanter}.  The Andreev reflection process, where an electron in N
hitting the NS interface is retro-reflected as a hole while a Cooper pair
is absorbed in S, yields transport by carriers of effective charge
$e^{*}\!=\!2e$.  As a result shot noise in an NS contact can be doubled
compared to the N case ($e^{*}\!=\!e$).  The first calculations were
performed within the quantum coherent transport theory
\cite{dejong,muzyk,martin}.  Experiments confirmed the doubling prediction
by direct low frequency current noise measurements \cite{nature} and
photon-assisted noise measurements in the GHz range \cite{yale}.  No
calculation was available at finite temperature T and voltage V, and the
role of phase coherence remained questionnable because, though it was
assumed by the theory, doubling was observed on a much larger bias range
\cite{nature}.  Nagaev and B\"uttiker recently developed a semiclassical
approach to noise in diffusive NS junctions which addresses these issues
\cite{nabu}.

Assuming that transport through the NS interface involves only Andreev
reflection below the superconducting gap $\Delta$ and only quasiparticles
of charge $e$ above $\Delta$, the authors determine the non-equilibrium
distribution function of electrons in the normal microbridge and calculate
the shot noise $S_{I}$ using the Boltzmann-Langevin method.  They obtain
the following analytical expression for $S_{I}(V)$ valid at finite V and
T, with $k$ the Boltzmann constant and $R$ the resistance:\cite{nabu}
\mybeginwide
$$
 S_I
 =
 4\frac{kT}{R}
 \left\{
   \frac{2}{3}
   +
   \frac{1}{3}
   \frac{eV}{kT}
   \coth
   \left(
     \frac{eV}{kT}
   \right)
   +
   \frac{1}{6}
   \left[
     \tanh
     \left(
       \frac{\Delta + eV}{2kT}
     \right)
     +
     \tanh
     \left(
       \frac{\Delta - eV}{2kT}
     \right)
     -
     2\tanh
     \left(
       \frac{\Delta}{2kT}
     \right)
   \right]
 \right.
$$
\begin{equation}
 \left.
   +
   \frac{1}{6}
   \left[
     \coth
     \left(
       \frac{eV}{2kT}
     \right)
     -
     2\coth
     \left(
       \frac{eV}{kT}
     \right)
   \right]
   \,
   \ln
   \left[
     \frac{
       \exp(\Delta/kT) + \exp(eV/kT)
     }{
       \exp(\Delta/kT) + \exp(-eV/kT)
     }
   \right]
 \right\}.
\label{S_I}
\end{equation}
\myendwide

This expression is plotted in figure \ref{data1k} (solid line) for
$T\!=\!1.35\,K$, $\Delta\!=\!1.2\,meV$ and the experimental R(V) values
(the non-linearity of R(V) on this voltage range equals 10{\%}).  Above the
low voltage regime of thermal noise ($eV\!\ll\!  kT$) the shot noise
increases linearly with current with a slope of $2\!\times\!\frac{2eI}{3}$,
twice the value for a normal contact.  This doubling is now predicted up to
$\Delta$, above which shot noise recovers the normal $\frac{2eI}{3}$ value. 
The $\frac{1}{3}$ factor comes from the diffusive nature of the normal
metal\cite{blanter}.  The dashed line in fig.\ref{data1k} simulates a
continuation of the doubled shot noise regime above the gap.  We previously
analyzed our results in Cu-Nb (NS) junctions by simply replacing the charge
$e$ by $2e$ in the analytical expression for a diffusive normal contact. 
This led to an excellent agreement up to $V\!\approx\!\Delta$, but could
not of course account for the recovery of the noise in the normal case
above the gap.  We also tried to apply the theory developped by Khlus
\cite{khlus} for ballistic NS constrictions at finite voltage, and
found only a qualitative rough agreement \cite{nature}.  This was expected
in absence of a more appropriate model.

\begin{figure}
\vspace{3mm}
 \centerline{
   \epsfxsize8cm
   \epsffile{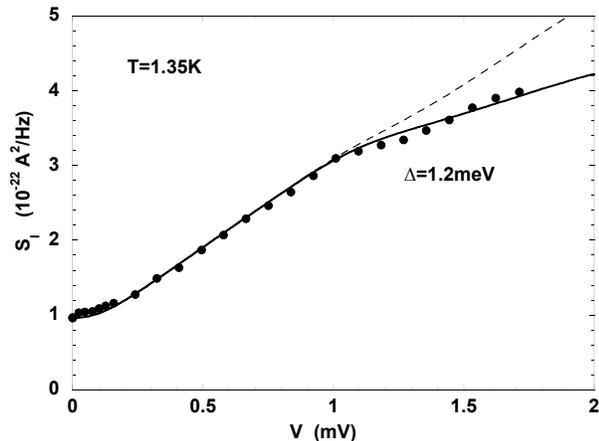}}
\caption{Shot noise measurements (dots) in an NS (Cu-Nb) junction compared
to the predictions (solid line) from the semiclassical theory \cite{nabu}
with the superconducting gap $\Delta$ as only parameter.  For
$V\!<\!\Delta$ the predicted doubled shot noise is confirmed
experimentally, as already described
\cite{nature}.  The dashed line simulates a doubled shot noise above 
$\Delta$ to quantitatively emphasize the difference with the normal case. 
For $V\!>\!\Delta$ an excellent quantitative agreement with the theory is
found for the first time.}
\label{data1k}
\end{figure}

Equation \ref{S_I} can be compared to experimental data as shown in
fig.\ref{data1k}, with $\Delta$ as only adjustable parameter.  The best fit
to our data at $1.35\,K$ is obtained with $\Delta=1.2\,meV$, a number very
close to the value for bulk Nb ($1.35\,meV$).  The perfect doubling of shot
noise was already observed up to $\Delta$, i.e. on a range where we showed
using reentrance measurements that phase coherence was absent\cite{nature}. 
The semiclassical prediction reinforces this observation, establishing that
doubling of shot noise requires essentially elastic interactions, but not
phase conservation.  For $V>\Delta$ an excellent agreement with the
semiclassical theory is also found.  This high bias region corresponds to
the case of a normal junction where transport occurs by carriers of charge
$e$.

At $4.2\,K$, equation \ref{S_I} with $\Delta =0.7\,meV$ yields a good
agreement with the data, as shown in figure \ref{data4k}.  However, because
$\Delta\approx 2kT$ at $4.2\,K$, electron-like excitations might be
involved in the transport, violating the basic assumption that only Andreev
reflections can account for the transport at $V<\Delta$.  In that case one
expects an effective charge between $e$ and $2e$, and a noise between the
doubled and normal case, in accordance with the data \cite{nature}.  A
quantitative comparison between the data at $4.2\,K$ and the semiclassical
theory cannot yield definitive conclusions because it takes place at this
high temperature regime where $kT\approx\Delta$.

\begin{figure}
\vspace{3mm}
 \centerline{
   \epsfxsize8cm
   \epsffile{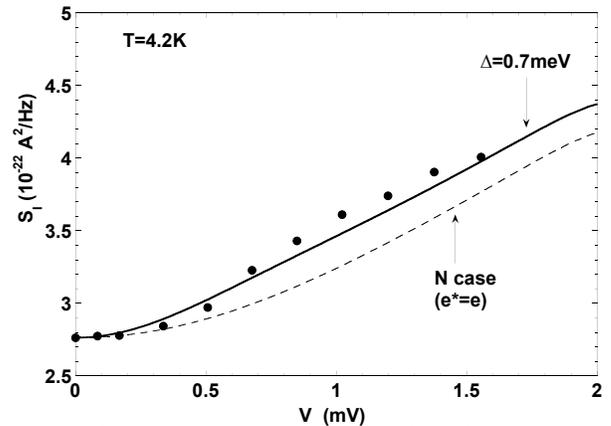}}
\caption{Shot noise at $4.2\,K$ compared to the semiclassical theory with 
$\Delta\!=\!0.7\,meV$ (solid line), and to the noise of a normal contact 
(dashed line). The measured noise lies in between the normal and doubled 
values, as qualitatively expected in the high temperature regime where $kT$ 
is not negligible compared to $\Delta$.}
\label{data4k}
\end{figure}

In conclusion we used the semiclassical theory presented by Nagaev and
B\"uttiker to quantitatively explain our shot noise data on diffusive NS
junctions at $1.35\,K$ over the whole bias range.  The excellent agreement
for doubled shot noise up to the gap confirms that phase coherence is not
necessary to observe the effect.  Above the gap the shot noise is found to
recover the value for a normal contact with a voltage-independent excess
noise, in quantitative agreement with the theory.

% ////////////////////  BIBLIOGRAPHY   \\\\\\\\\\\\\\\\\\\\

\end{multicols}

\end{document}